\documentclass[epj]{svjour}
\usepackage{graphicx}
\usepackage{amssymb}

\newcommand{\gev}{\,{\rm GeV}}
\newcommand{\mev}{\,{\rm MeV}}
\newcommand{\mpi}{m_\pi}

\begin{document}
\title{Strangeness contributions to nucleon form factors}
\author{Ross D.~Young
}                     
\institute{Jefferson Lab, 12000 Jefferson Ave.,
           Newport News, VA 23606 USA}
\date{Received: date / Revised version: date}
%
\abstract{
We review a recent theoretical determination of the strange quark
content of the electromagnetic form factors of the nucleon. These are
compared with a global analysis of current experimental measurements
in parity-violating electron scattering.
\PACS{
      {14.20.Dh}{Protons and neutrons}   \and
      {11.30.Er}{Charge conjugation, parity, time reversal, and other discrete symmetries in particles and fields}
     } 
} 
\maketitle
\section{Introduction}
The determination of the strange quark content of the nucleon offers a
unique probe to measure the nonperturbative structure of the
nucleon. As the nucleon carries zero net strangeness, the influence of
strange quarks arises entirely through interaction with the
vacuum. Technically speaking, strange quarks directly probe the role
of the fermion determinant in QCD.  While strangeness measurements in
nucleon structure have been difficult to isolate, the contribution of
the neutral weak current in elastic scattering offers perhaps the most
direct measurement of the strange quark content of the
nucleon \cite{PVES}.

Here we review recent progress in the study of the strange quark
contributions to the nucleon form factors. In Section \ref{sec:chiral}
we discuss the theoretical developments in the chiral extrapolation of
lattice simulation results that have enabled a precise determination
of the strangeness form factors. An outline of this determination is
provided in Section \ref{sec:thy}. In Section \ref{sec:exp} this
theoretical prediction is compared with a global analysis of the
experimental measurements searching for strangeness in the nucleon.

\section{Chiral applications in lattice QCD}
\label{sec:chiral}
The computational expense of incorporating the effects of the fermion
determinant has restricted modern lattice QCD simulations to the use
pion masses that are typically $m_\pi\gtrsim 500\mev$. Recent progress
has seen nucleon 3-point functions simulated with pion masses pushing
down to the $350\mev$ range
\cite{Edwards:2005ym,Alexandrou:2006ru,Gockeler:2006ui}, yet a
reliable extrapolation in the pion mass is still required to compare
with reality --- until the physical point is readily accessible.

Ultimately, chiral perturbation theory offers the potential to deliver
model-independent quark-mass extrapolations of lattice results. As
disappointing as it may be, there is mounting evidence that
applications of low-order chiral expansions should be taken with
serious caution beyond pion masses of the order $300\mev$
\cite{Young:2002ib,converg}.
Further, the situation could be significantly worse for observables
which are particularly singular near the chiral limit, such as
magnetic moments \cite{Young:2004tb}, charge radii or
polarisabilities.

In the future, chiral extrapolations will be constrained {\em
model-independently} by precision, large-volume lattice calculations in the chiral
regime. Until then, one requires methods which
can reliably extrapolate from the moderately-heavy quark mass regime,
while maintaining all the constraints of the effective
field theory. The best available solution is to
reformulate the effective field theory using finite-range
regularisation (FRR) \cite{Young:2002ib}. 

In extrapolating lattice simulation results from beyond the chiral
regime, one cannot guarantee that results will be independent of
regularisation scheme. By choosing a particular scheme, one has
necessarily introduced a model --- whether it be FRR or a more
traditional regularisation. The advantages of FRR have been
quantitatively demonstrated for the nucleon mass. Using lattice
results over the range $0.25\gtrsim\mpi^2\gtrsim 1.0\gev^2$, the FRR
extrapolated nucleon mass at the physical point displays less than 1\%
variation associated with the truncation between successive orders in
the chiral expansion. Further, the sensitivity to the choice of
functional form of FRR is also less than 1\%
\cite{Leinweber:2003dg}. Until sufficient lattice results are
available in the chiral regime, when the choice of regularisation
becomes superfluous, FRR offers {\em independent-of-model} chiral
extrapolations.

Because of the cost of simulating the fermion determinant,
historically it has been common in lattice QCD to ignore this
contribution to the path integral. This is the quenched
``approximation'', where the influence of quark-antiquark
pair-creation in the vacuum is neglected.  Fortunately, the study of
the chiral extrapolation of baryon masses in quenched and dynamical
simulations has revealed a remarkable phenomenological relation
between these simulations.  The differences between quenched and
dynamical baryon masses are well described by the differences in the
Goldstone boson loop corrections of the low-energy effective field,
when evaluated with an appropriate finite-range regulator
\cite{Young:2002cj}.  Although this is not a field-theoretic
connection, the numerical success does mean that one has substantial
confidence in obtaining physical estimates from quenched lattice
results.

Beyond the baryon masses, the technique of chiral unquenching has been
extended to the nucleon magnetic moments \cite{Young:2004tb}. Here it
was predicted that there should be very little difference in the
quenched and dynamical nucleon magnetic moments over a large range of
quark masses, with significant differences only anticipated near the
chiral limit.  These findings have been recently supported by first
calculations with 2-flavour dynamical lattices
\cite{Alexandrou:2006ru,Gockeler:2006ui}.

With the success of chiral extrapolations and the estimation the
effects of the quark determinant, we look to the extraction at the
strangeness contributions to the nucleon electromagnetic form factors.

\section{Strangeness calculation}
\label{sec:thy}
Direct lattice QCD calculations of the strangeness content have 
been unable to produce a conclusive determination
\cite{strangeLat}.  It is hoped that the
next generation of calculations could shed light on this elusive
signal. This may require further development of emerging lattice
techniques. One potential gain could be seen by utilising background
field methods \cite{bgf}, where a weak signal
could be enhanced by coupling a strong electromagnetic field to the
vacuum strange quarks. The method to evaluate the all-to-all
propagator developed by the Dublin group offers significantly improved
precision over traditional stochastic estimators \cite{Foley:2005ac},
and it would be interesting to see this applied to a strangeness form
factor calculation.

\begin{figure}[t]
\begin{center}
{\includegraphics[height=0.4\columnwidth,angle=90]{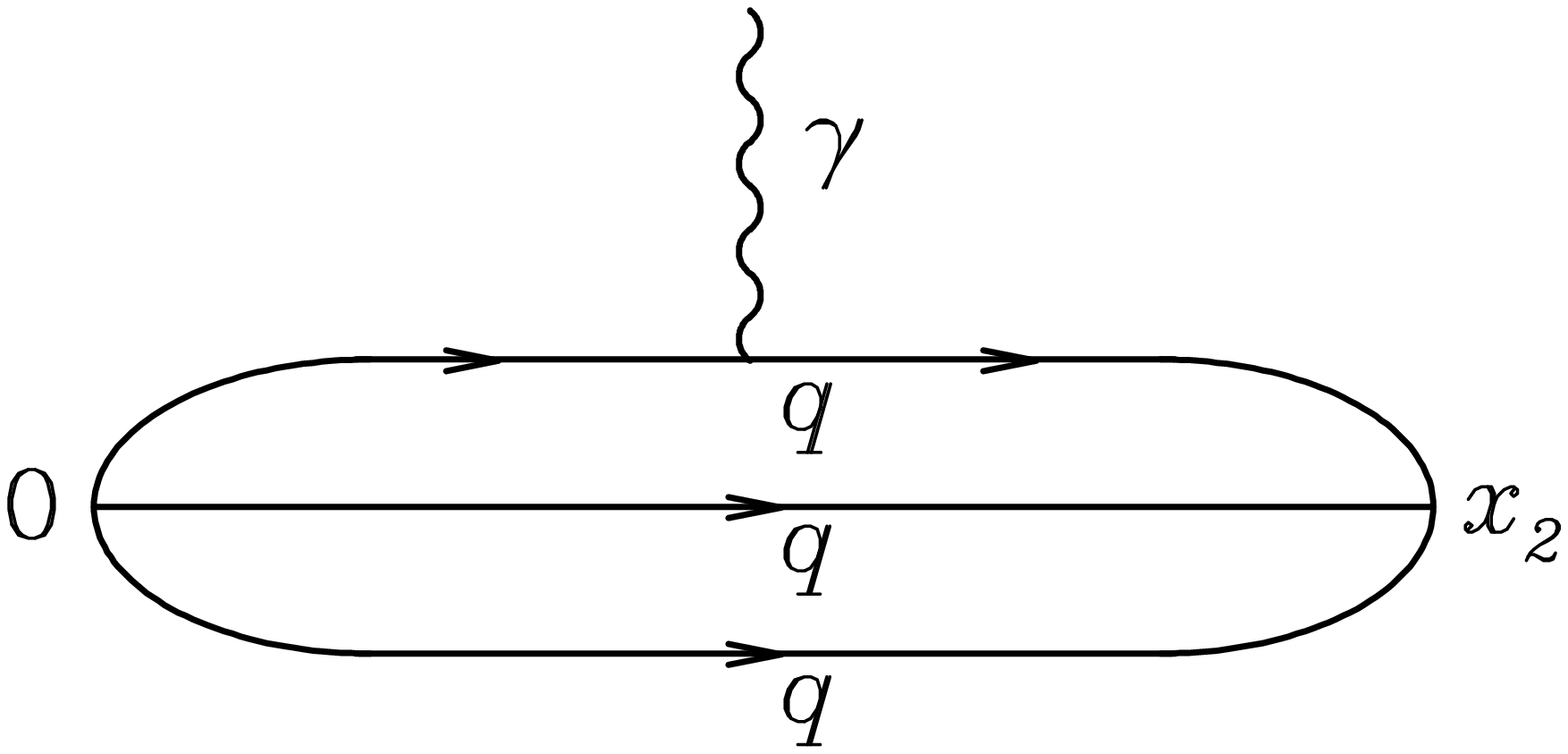}
\hspace*{5mm}
\includegraphics[height=0.4\columnwidth,angle=90]{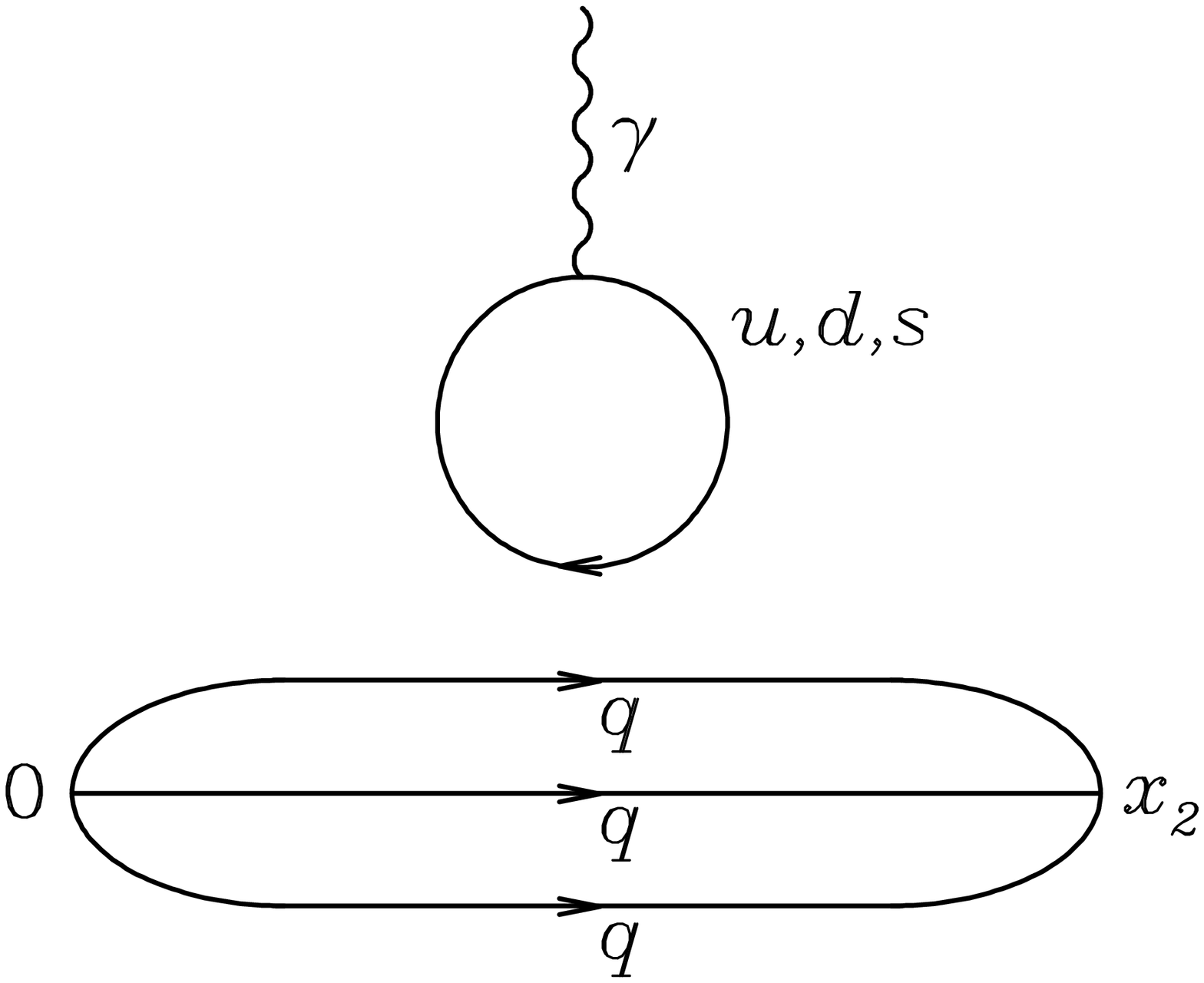}}
\end{center}
\caption{Connected (left) and disconnected (right) contributions to
baryon 3-point functions.}
\label{fig:top}
\end{figure}

While awaiting the development of these techniques, one must rely on
more indirect methods for an accurate extraction of the strangeness
form factors. By manipulating the QCD path integral, one can isolate
the various quark contributions to baryon 3-point functions
\cite{Leinweber:1995ie}. Using this decomposition, combined with
charge symmetry \cite{chargeSym} and the experimentally measured
hyperon radii, the sensitivity to lattice systematics can be
significantly reduced \cite{Leinweber:1999nf}. The strangeness
magnetic moment can be written as
\begin{equation}
G_M^s = \frac{{}^lR^s_d}{1-{}^lR^s_d}\left[ 2p+n -
  \frac{u^p}{u^\Sigma}(\Sigma^+-\Sigma^-)\right]\,,
\label{eq:doub}
\end{equation}
\begin{equation}
G_M^s = \frac{{}^lR^s_d}{1-{}^lR^s_d}\left[ p+2n -
  \frac{u^n}{u^\Xi}(\Xi^0-\Xi^-)\right]\,,
\label{eq:sing}
\end{equation}
where $p$, $n$, $\Sigma^\pm$ and $\Xi^{0/-}$ denote the experimentally
measured magnetic moments of the respective baryon.  The formulae each
rely on two inputs from lattice simulations.  The first is the ratio
$u^p/u^\Sigma$, which measures the relative strength of the valence
(Fig.~\ref{fig:top}a) $u$-quark contribution in the proton relative to
the $\Sigma^+$ --- or similarly $u^n/u^\Xi$ in Eq.~(\ref{eq:sing}).
The second is ${}^lR^s_d$, which describes the ratio of the
strange-to-light disconnected (Fig.~\ref{fig:top}b) contributions.

\begin{figure}[tbp]
\begin{center}
{\includegraphics[height=\columnwidth,angle=90]{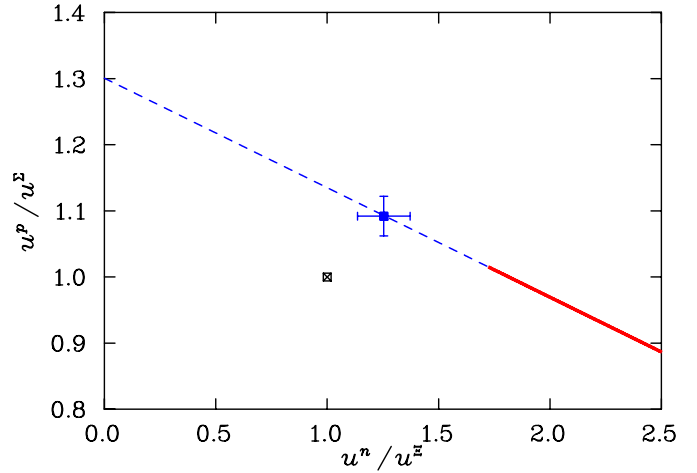}}
\end{center}
\caption{The line (dashed $G_M^s(0) < 0$, solid $G_M^s(0) > 0$)
indicates the charge symmetry constraint on the ratios
$u^p/u^{\Sigma}$ and $u^n/u^{\Xi}$.  The crossed square indicates the
point corresponding to environment independent quark moments.  Our
determination is illustrated by the filled square
\cite{Leinweber:2004tc}.  }
\label{fig:SelfCons}
\end{figure}

Equating equations (\ref{eq:doub}) and (\ref{eq:sing}) and using the
experimental magnetic moments produces a linear relationship between
the two unknown valence ratios. This constraint, a result of charge
symmetry alone, is displayed in Figure~\ref{fig:SelfCons}.
The line is divided by two segments, where the sign of $G_M^s$ can be
determined under the quite general assumption that
$0>{}^lR^s_d>1$. Recently it has been suggested that there could be a
sign change in this ratio between the heavy-quark limit and naive
expectations in the Goldstone boson sector \cite{Ji:2006vx}. Given
that the properties of the kaon are much more Goldstone-like than a
heavy-light meson, and that the heavy-quark limit of $\mu_p/\mu_n$ is
approached very slowly \cite{Leinweber:2001ui}, it should be not be
expected that the strange quark could be reliably described by
heavy-quark effective theory.

The techniques discussed in Section~\ref{sec:chiral} were applied to
determine the ratios $u^p/u^{\Sigma}$ and $u^n/u^{\Xi}$, appearing in
Eqs.~(\ref{eq:doub}) and (\ref{eq:sing}). The analysis has utilised a
high-precision numerical study of the baryon electromagnetic form
factors in quenched lattice QCD \cite{Boinepalli:2006xd}. Upon
performing finite-volume corrections, adjustments for the quenched
approximation and a controlled chiral extrapolation, the resulting
ratios are compared with the experimental constraint in
Figure~\ref{fig:SelfCons} \cite{Leinweber:2004tc}. The excellent
agreement with the constraint from experiment is a first check on the
consistency of our calculation.

\begin{figure}[tbp]
\begin{center}
{\includegraphics[height=\columnwidth,angle=90]{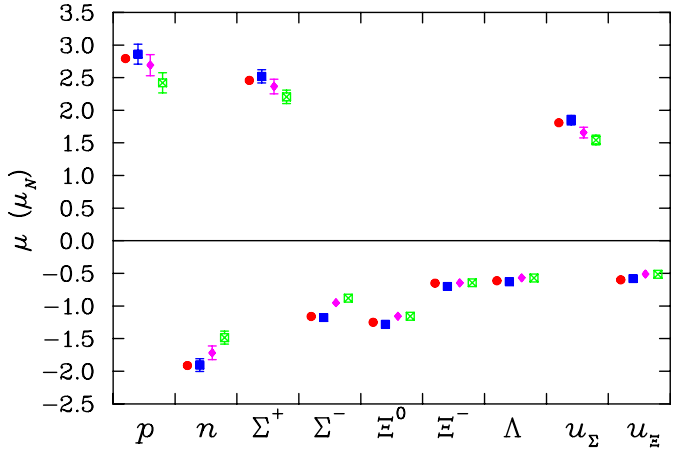}}
\end{center}
\caption{The extracted full-QCD magnetic moments ($\blacksquare$)
\cite{Leinweber:2004tc} display excellent agreement with the
experimental moments ($\bullet$). To indicate the size of corrections,
the quenched ($\blacklozenge$) and finite-volume quenched
($\boxtimes$) are also displayed.}
\label{fig:MagMom}
\end{figure}

Using the same procedure which enabled the correction from quenched
to dynamical within the valence sector, we have also estimated the
disconnected contributions to reconstruct the full magnetic moments in
QCD. In Figure \ref{fig:MagMom} the full magnetic moments are shown
with the experimentally measured values for the entire baryon
octet. The agreement with experiment is remarkable, offering further
support for the validity of this analysis. Further, Figure
\ref{fig:MagMom} also displays our excellent reproduction of the two
experimental valence moments, $u^\Sigma$ and $u^\Xi$.

With the valence ratios determined, the final input required from
Eqs.~(\ref{eq:doub}) and (\ref{eq:sing}) is the ratio
${}^lR^s_d$. As discussed above, there have been technical
difficulties in extracting a signal for the disconnected insertion
directly. Therefore we have used the relative magnitude of the
strange-to-light disconnected insertions, estimated through the same
method that applied the unquenching corrections and constructed the
full moments displayed in Figure \ref{fig:MagMom}. Given that the
magnitude of the valence sector is so well reproduced, particularly by
$u^\Sigma$ and $u^\Xi$, and that the overall scale of the disconnected
insertions is observed to correctly adjust the full moments from the
valence-only sector, we have substantial confidence that our estimate
is accurate.

The final result for the strange magnetic moment of the proton is
$G_M^s = -0.046 \pm 0.022 \mu_N$ \cite{Leinweber:2004tc}, an error of
just two hundredths of a nuclear magneton. The error quoted includes
the best possible estimates of the statistical and systematic errors
in the calculation \cite{Leinweber:2005bz}. The announcement by that
G0 Collaboration that $G_M^s$ was positive, at the 95\% confidence
level \cite{Armstrong:2005hs}, posed a significant challenge to
this theoretical result.

Recently, the same techniques have been applied to theoretically
extract the strange electric form factor of the
proton~\cite{Leinweber:2006ug}. The absence of accurate data for octet
charge radii meant that it was necessary to use absolute values of the
unquenched contributions of the $u$ and $d$ quarks to the charge
radius of the proton. Again the result was particularly accurate, with
$G_E^s(Q^2 = 0.1\gev^2) = +0.001 \pm 0.004 \pm 0.004$. This is
in quite good agreement with the published measurement by 
the HAPPEx Collaboration at Jefferson Lab, 
$G_E^s(Q^2 = 0.1\gev^2) = -0.01 \pm 0.03$
\cite{HAPPEX}.

With the increasing collection of strangeness measurements, and the
promise of even more accurate data from HAPPEx in 2006, it is
especially timely to see consolidated treatment of the strange form
factor extraction. Recently, a systematic analysis of the published
world data has been performed in Ref.~\cite{Young:2006jc}.  The next
section briefly summarizes the input to that analysis and its main
conclusions.

\section{Global analysis of experiment}
\label{sec:exp}
Since the first results reported by the SAMPLE Collaboration in 1997
\cite{Mueller:1997mt}, many measurements of the parity-violating
contribution to the elastic form factors of the nucleons have now been
completed. These experiments have been performed on several targets,
the proton, deuteron and helium-4, and at various kinematic
configurations. The measured parity-violating asymmetries are
sensitive to the strange electric and magnetic form factors in
different linear combinations. By combining the results of these
measurements, one can separate the electric and magnetic
contributions.

In addition to probing the strange vector current, the experimental
asymmetries are also sensitive to the the weak axial current in the
nucleon, which become increasingly more significant at backward
scattering angles. The axial-$Z$ coupling to the nucleon can be
controlled through semileptonic decays, deep-inelastic scattering and
careful treatment of radiative corrections.  In addition, there is a
parity-violating photon coupling, which is sensitive to an odd-parity
component of the nucleon wavefunction --- the anapole form factor
\cite{Haxton:1989ap}. Being nonperturbative in origin, one is forced
to introduce another unknown form factor which must be determined by
data. Because a single stand-alone experiment is sensitive to several
unknown form factors, the strangeness content has previously been
reported by imposing theoretical estimates for the anapole
contributions \cite{Zhu:2000gn}. With the extensive experimental
programs now completed, a global fit of all the data permits the
extraction of all unknown form factors \cite{Young:2006jc}, without
any need for theoretical input.

The parity-violating asymmetries have been constructed with a
consistent set of inputs to avoid introducing any systematic
distortion in the extracted form factors. In particular, the
asymmetries have all be computed using the same set of electromagnetic
form factors \cite{Kelly:2004hm} and the latest radiative corrections
\cite{Eidelman:2004wy}.

The cleanest separation of the unknown form factors is available at
$Q^2\sim 0.1\gev^2$, where the most substantial coverage of
measurements have been performed \cite{HAPPEX,exp}. With further data collected
in the near $Q^2$-vicinity, particularly by the G0 Collaboration
\cite{Armstrong:2005hs}, it is useful to use as much of the available
data as possible. Using measurements over a range of $Q^2$ values
necessitates introducing a parameterisation of the $Q^2$ evolution of
the form factors.  A Taylor expansion of the strange electric and
magnetic form factors has been utilised, defining
\begin{equation}
G_E^s = \rho_s Q^2 +\rho_s' Q^4 +\ldots\,, \qquad
G_M^s = \mu_s + \mu_s' Q^2 + \ldots\,.
\end{equation}
Provided the range of $Q^2$ values does not extend too
high, this gives a systematic technique to combine a large set of
data.

Because anapole contribution contributes together with the asymmetry arising
from the axial charges, for simplicity the same dipole form is
chosen for the axial and anapole contributions, with
\begin{equation}
\tilde{G}_A^N=\tilde{g}_A^N(1+Q^2/M_A^2)^{-2}\,,
\end{equation}
and
\begin{equation}
\tilde{g}_A^N = \left( \xi_A^{T=1} g_A \tau_3 + \xi_A^{T=0} a_8 + \xi_A^0 a_s \right) + \left( A_{\rm ana}^{T=1} \tau_3 + A_{\rm ana}^{T=0} \right)\,.
\end{equation}
The factors, $\xi$, denote the tree-level plus radiative corrections
multiplying the various axial charges. These charges are relatively
well known, with $g_A=1.2695$, $a_8=0.58\pm0.03\pm0.12$ and
$a_s=-0.07\pm0.04\mp0.05$. The second error in $a_8$ and $a_s$ denotes
a correlated uncertainty allowing for potential violations of
SU(3)-flavour symmetry in semileptonic hyperon decay. The second
bracketed term describes the anapole form factor, for which there is
only limited phenomenological information. Zhu {\it
et~al.}~\cite{Zhu:2000gn} have estimated the magnitude and Maekawa
{\it et~al.}~\cite{Maekawa:2000bd} the leading momentum
dependence. The approach taken here, is to determine this contribution
from the data, free from theory input. Given the size of the
uncertainties of the charges, $\tilde{g}_A^p$ and $\tilde{g}_A^n$, the
error is dominated by the poor knowledge of the anapole
contributions.

\begin{figure}[tbp]
\begin{center}
{\includegraphics[width=\columnwidth]{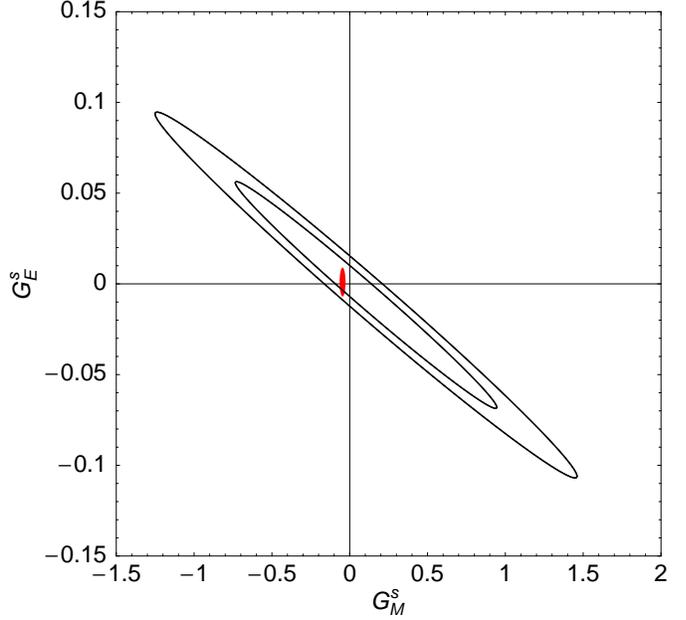}}
\end{center}
\caption{Determination of the strangeness magnetic and electric form
factors at $Q^2=0.1\gev^2$, 68\% and 95\% confidence intervals are
shown by the contours. The solid ellipse depicts the theory
result described in Section~\ref{sec:thy} \cite{Leinweber:2004tc,Leinweber:2006ug}.}
\label{fig:GMsGEs}
\end{figure}

The fit to the complete set of world PVES data at $Q^2<0.3\gev^2$
 yields best-fit parameters
\begin{eqnarray}
\tilde{g}_A^p &=&  0.05 \pm 1.38 \mp 0.29 \,, \\
\tilde{g}_A^n &=&  2.61 \pm 2.27 \mp 0.37 \,, \\
\rho_s &=&        -0.06 \pm 0.41 \mp 0.00 \gev^{-2} \,,\\
\mu_s &=&          0.12 \pm 0.55 \pm 0.07 \,,
\end{eqnarray}
where the first error denotes the uncorrelated experimental
uncertainty and the second the correlated uncertainty in the G0
experiment. The joint determination of the strangeness electric and
magnetic form factors at $Q^2=0.1\gev^2$ is shown in
Figure~\ref{fig:GMsGEs}, where we also compare with the theoretical
prediction described above.

\begin{figure}[tbp]
\begin{center}
{\includegraphics[width=\columnwidth]{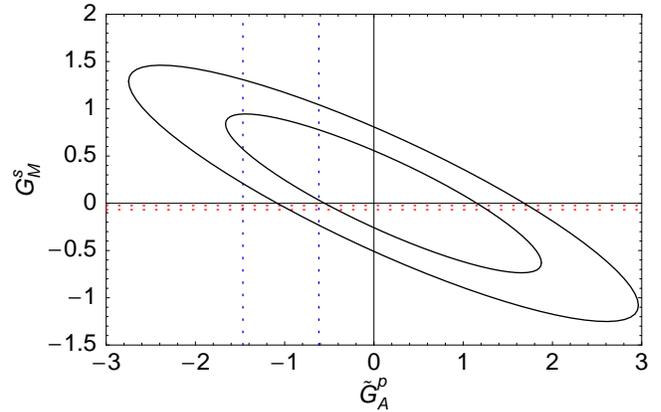}}
\end{center}
\caption{The contours display the 68\% and 95\% confidence intervals
for the joint determination of $\tilde{G}_A^p$ and $G_M^s$ at
$Q^2=0.1\gev^2$.  The horizontal and vertical bands display
the theory results of Leinweber et al.~\cite{Leinweber:2004tc} and Zhu
{\it et~al.}~\cite{Zhu:2000gn}, respectively.}
\label{fig:GAGM}
\end{figure}

\begin{figure}[tbp]
\begin{center}
{\includegraphics[width=\columnwidth]{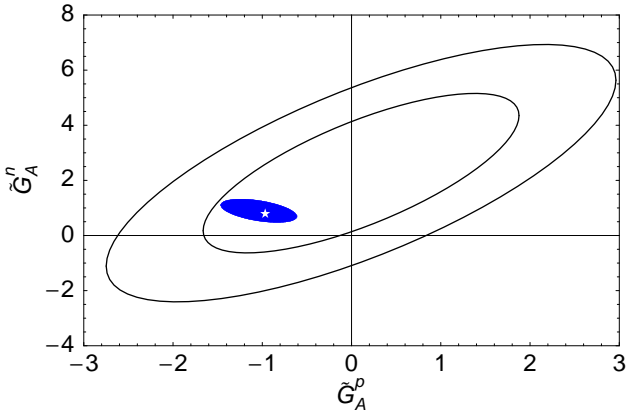}}
\end{center}
\caption{The contours display the 68\% and 95\% confidence intervals
for the joint determination of $\tilde{G}_A^p$ and $\tilde{G}_A^n$ at
$Q^2=0.1\gev^2$. The disc depicts the result of Zhu {\it
et~al.}~\cite{Zhu:2000gn}, with the white star indicating a null anapole form factor.}
\label{fig:Axial}
\end{figure}

The 68\% and 95\% confidence intervals for the determination of
$G_M^s$ against $\tilde{G}_A^p$ is shown in Figure~\ref{fig:GAGM} and
$\tilde{G}_A^n$--$\tilde{G}_A^p$ in Figure~\ref{fig:Axial}.
This helps provide some picture as to the four-dimensional parameter
space that is being constrained by the data. Within the experimental
uncertainty on the determination of the axial form factors,
Figure~\ref{fig:GMsGEs} shows the strangeness form factors mapping out
a long, yet narrow, region of parameter space --- including the point
of vanishing strangeness. Going to the space of the axial form
factors, within the strangeness determination, these are quite poorly
constrained in comparison to the theoretical result of Zhu {\it
et~al.}~\cite{Zhu:2000gn}. Nevertheless, they are totally consistent
with this calculation, which includes the point of vanishing anapole
form factor.

Independently of each other, the strangeness and anapole contributions
appear to be consistent with zero, and hence in agreement with both
the theoretical results in question.
Figure~\ref{fig:GAGM} indicates that the two theory results appear to
be outside the 68\% confidence level to be simultaneously supported by
the data.  In the complete four-dimensional space, it is found that
there is a 92\% support for a nonzero value in at least one of the
strange or anapole form factors \cite{Young:2006jc}.

\section{Summary}
A series of developments in the study of the chiral extrapolation
problem in lattice QCD, combined with a high-precision numerical
calculation of baryon electromagnetic form factors, has enabled a
detailed study of the strangeness content of the nucleon.  A precise,
small negative value has been predicted for the strangeness magnetic
moment. The strangeness electric contribution is found to be bounded
within half a percent of the total mean-square charge radius of the
proton.

The determined strangeness form factors are found to be in good
agreement with a global analysis of the world strangeness
measurements.  The anapole contributions in parity-violating elastic
scattering are also consistent with modern theoretical estimates ---
although there is a small hint that both theory results cannot both be
supported by the experimental data simultaneously. We look forward to
future measurements which will further expand our view of the flavour
structure of the nucleon.

\section*{Acknowledgements}
Thanks are extended to the many collaborators who have been involved
in the research presented here, particularly D.~Leinweber and
A.~Thomas. This work was supported by U.S. DOE Contract
No. DE-AC05-06OR23177, under which Jefferson Science Associates, LLC
operate Jefferson Lab.

\end{document}